\documentclass[11pt]{article}
\usepackage[utf8]{inputenc}
\usepackage{amsmath, amsfonts, amssymb}

\usepackage[a4paper, margin=1in]{geometry}
\usepackage{setspace}
\usepackage{newpxtext}
\usepackage{authblk}
\usepackage{graphicx}

\usepackage[hyperindex,breaklinks]{hyperref}
\usepackage{xcolor}
\usepackage{booktabs}
\usepackage[group-separator={,}]{siunitx}
\usepackage{float}
\usepackage{wrapfig}

\title{The Paradox of Collective Certainty in Science}

\author[1]{Eamon Duede\thanks{Email: eduede@g.harvard.edu}}
\author[2,3,4]{James Evans\thanks{Email: jevans@uchicago.edu}}
\affil[1]{Harvard University}
\affil[2]{University of Chicago}
\affil[3]{Santa Fe Institute}
\affil[4]{Knowledge Lab}

\date{} 

\begin{document}

\maketitle

\begin{abstract}
   We explore a paradox of collective action and certainty in science wherein the more scientists research together, the less that work contributes to the value of their collective certainty. When scientists address similar problems and share data, methods, and collaborators, their understanding of and trust in their colleagues' research rises, a quality required for scientific advance. This increases the positive reinforcement scientists receive for shared beliefs as they become more dependent on their colleagues' knowledge, interests, and findings. This collective action increases the potential for scientists to reside in epistemic ``bubbles'' that limit their capacity to make new discoveries or have their discoveries generalize. In short, as scientists grow closer, their experience of scientific validity rises as the likelihood of genuine replication falls, creating a trade-off between certainty and truth.
\end{abstract}

\section*{The Social Limits of Scientific Certainty}

In this chapter, we develop and explore a perplexing epistemological outcome of social dynamics in science, which involves a paradox of collective action and certainty: The more scientists work together in research, the less their collaborative work contributes to the value of collective certainty \cite{belikov2022prediction, danchev2019centralized}. This results from the following socio-cognitive process. As scientists come to address similar problems and share data, methods, and collaborators to that end, their understanding of and trust in their colleagues' research rises. Ironically, however, this increases the positive reinforcement they receive for shared beliefs as they become less independent of those colleagues, their interests, and findings \cite{kang2023limited,sourati2022data}. This raises the likelihood that scientists come to reside in an epistemic ``bubble'', neither replicable nor robust to perturbations in research design or application. In short, as scientists grow closer, the experience of scientific validity rises as the likelihood of replication falls, creating a trade-off between certainty and truth.

The paradox associated with this situation is that working together is a prerequisite for epistemic trust. If one scientist cannot understand a second's methods and materials, they have no direct justification for their findings. If the first does not have access to the reputation of the second or the provenance of their data and approach, then they have no indirect justification. Without either, there is no basis for trust \cite{shapin1995social,fricker2006second,frost2013moral,gerken2015epistemic}. 

We note that these same dynamics are equally present but somewhat less paradoxical for the process of collective discovery. Diverse and disconnected perspectives have been shown to systematically generate research surprises and their resolution, or the identification of novel patterns \cite{shi2023surprising, duede2021social}. The connection of diverse perspectives in discovery is commonly associated with interactions that feel serendipitous \cite{duede2024being}. This becomes paradoxical in the context of replication because researchers rarely \textit{accidentally} confirm a scientific finding. Rather, they purposively \textit{intend} to confirm it because they are scientifically proximate, which decreases the independence and robustness of their confirmations.   

In this paper, we first rehearse contemporary concerns over a replication crisis. Then, we articulate and illustrate this collective epistemic puzzle more precisely. Finally, we explore a range of possible institutional solutions to the paradox.

\section*{The ``Replication Crisis'' in Science}

The central output of the modern scientific process is a distribution of overlapping and competing truth claims about natural, social, and artificial worlds, published in papers and preprints, proclaimed in face-to-face conferences, and muttered in conversation. A stated motivation for science is the organized identification and dissemination of scientific facts. Nevertheless, from historical data on shifts in scientific methodology \cite{Henrion1986-an,shapin2011leviathan} and epistemological standards \cite{stegenga2023justifying, Knorr-Cetina2013-cr, cartwright1984laws, strevens2020knowledge}, we know that the correlation between facts and published claims lies far from identity. The \textit{reproducibility} of a research finding refers to the scientific community's ability to obtain the same results using the same methods and data. The \textit{replicability} of such a study references the community's ability to achieve consistent results using the same methods but different data. And the \textit{conceptual replicability} of a study suggests the community's ability to achieve the same results using different methods and different data but the same `concept'. Conceptually replicable results are the most robust \cite{wimsatt1981robustness} and are what we, perhaps pre-reflectively think of when we think of `valid' scientific claims. To simplify things a bit, in what follows, we will often use the terms `reproducibility' and `replicability' interchangeably to denote scientific results that are \textit{merely} replicable, \textit{seemingly} empirically robust, but, in fact, lack conceptual replicability or validity.

Over the past two decades, the non-reproducibility of many scientific claims has become more widely acknowledged \cite{Ioannidis2005-gx, Rzhetsky2006-oa}. Biomedical replication studies were inspired by early concerns about reproducibility and reported reproduction rates from preclinical peer-reviewed publications at between 10 and 25\% \cite{prinz2011believe,begley2012raise}. These and related demonstrations have led to an experienced and publicized `replication crisis' in a growing number of fields, especially genetics \cite{ioannidis2001replication, ioannidis2007non}, medicine \cite{ioannidis2013replicate}, and social psychology \cite{maxwell2015psychology, wiggins2019replication}, with many other fields following \cite{king1995replication, berry2017assessing}. Some areas of inquiry, such as behavioral priming research in psychology, have received sufficient skeptical scrutiny with the seemingly plausible outcome of reducing the number of such studies published in major outlets \cite{yong2012nobel, cesario2014priming}.  

Replication failures are widely attributed to systemic biases theorized as the misalignment of scientific incentives \cite{nissen2016publication, smaldino2016natural}. 
Most prominently, the scientific publication system rewards novel discoveries over confirmations \cite{Ioannidis2005-gx, nosek2015promoting, alberts2015self, errington2014open}, which leads to a preponderance of questionable research choices such as \textit{p}-hacking \cite{simonsohn2014p, head2015extent, brodeur2020methods}, 
`flexible' data analysis \cite{simmons2011false}, 
selective reporting (the ‘file drawer problem’) \cite{rosenthal1979file}, low statistical power \cite{dumas2017low}, and confirmation bias \cite{nuzzo2015fooling}. These questionable decisions, combined with incomplete reporting of statistical methods and data \cite{nosek2015promoting}, contribute to the publication of false results that are unlikely to replicate in future experiments \cite{simmons2011false}.

Concerns over the reproducibility and replicability of science raise concerns about the fundamentally cumulative process of building on prior published results. Furthermore, these concerns cast doubt on the possibility of peer review and the ability of experts to assess a scientific claim's validity from its detailed description alone. Consider what research reproductions or replications require in capital-intensive research fields, including genomics, analytical chemistry, and high-energy physics: expensive, specialized equipment, skilled laboratory technicians, routine checks that confirm sensitivity, and calibration to `de-bias' instruments relative to established baselines from the scientific and technical literature, or at the complexity limit (e.g., particle accelerators) based on detailed simulations of the instrument itself \cite{Knorr-Cetina2013-cr}. In turn, carrying out tasks of these kinds demands complex communication through the transfer of skilled bodies, scientific and technical documents, simulation and analysis code, and large quantities of data.

In these settings, the production of scientific results appears similar to the production of complex technologies, like semiconductor fabrication facilities \cite{monch2012production} or the TEA laser, which in 1975 had never been successfully assembled without people with prior experience in assembling it \cite{collins1975building, collins1992changing}. Nevertheless, the significance of tacit knowledge for science and tacit knowledge for technology is distinct. The construction of technology has required proof of reproduction, as demanded by patent offices since the eighteenth century, but not precise or conceptual replication. Why? Because if the technology's manufacture can reproduce its performance, then it may be irrelevant whether or not an alternative design exists that could accomplish the same task. Bruno Latour highlights this point in his analysis of Pasteur's demonstration of an anthrax cure in the French countryside \cite{latour1993pasteurization}. Latour's demonstration succeeded, first, by controlling anthrax within his laboratory, but second, by replicating his laboratory in the French countryside. Only farms that could maintain lab-like cleanliness were considered to demonstrate his vaccine. By contrast, for a complex scientific experiment that makes claims of generality beyond the context of its initial design and demonstration, inability to replicate implies failure. In sum, both scientific and technological reproduction rely on tacit knowledge, but this poses a special challenge for generalized scientific knowledge rather than for technological production.

\section*{The Dynamics of Collective Certainty}

The dynamics of collective scientific investigation give rise to a puzzle. For scientists to pursue scientifically important problems in their fields, they must first agree on and iteratively ratify investigative elements that constitute the epistemic norms of those fields. Without a consensus on markers of importance, acceptable data and methods, and justificatory standards required for the ratification or rejection of claims (e.g., ``causal identification'' in economics, ``performance relative to SOTA'' in computer science \cite{koch2021reduced}), scientists would be unable to evaluate the quality and value of emergent research. Absent familiar problems, approaches, literature, and researcher genealogies, nothing would signify membership in a field beyond signification itself.

To constitute a discipline, scientists must agree on the problems and approaches most likely to solve them. This process of agreement and the associated homogeneity of experience and perspective it engenders necessarily decreases the intellectual, methodological, and epistemological independence across the fields' investigations. The upshot is that scientists can select problems and build on solutions recognizable from the perspective of the field. They deploy standardized methods as appropriate for addressing shared problems. And they justify their results in accordance with established disciplinary standards, which continuously reconstitute them \cite{strevens2020knowledge}. Philosopher Thomas Kuhn believed that something akin to this process occurs in the iterative imitation of exemplars \cite{kuhn2012structure}. Sociologist Pierre Bourdieu argued for a complementary process with respect to the conventions of investigation within a field acquired through conventional education, mentorship, and practice as a scientific \textit{habitus} \cite{bourdieu2004science, bourdieu2020outline}.

Nevertheless, the history of science has repeatedly demonstrated that the selection of field representative exemplars carries epistemic risks for progress. Consider, for example, the selection of central methods. First-movers in a field are often those who develop a novel method to solve a particular problem (e.g., Richard Feynman with his physical diagrams for heuristic computing in quantum field theory and related physics \cite{kaiser2019drawing}). The epistemic utility of this method is then replicated by scientists in the field through successfully repeated cases involving the iterative reuse of the method. In general, however, a method is useful to the extent that cases for which it is applicable are well addressed by it. As we show below, in elevating and consolidating disciplinary efforts around particular methods, fields typically experience a reduction in marginal epistemic returns because scientists rationally selected the best cases first. Under this common scenario, cases for which agreed-upon methods are optimal become rapidly exhausted.

For example, early work on symbolic systems dominated artificial intelligence research for decades, scoring early successes that crowded out interest in and funding for connectionist methodologies \cite{dreyfus2012making}. Yet, these early successes had subsequent difficulty generalizing beyond a limited set of problem settings characterized by well-defined rules and precise objectives (e.g., first-order logic used for chess play and theorem proving). The mid-century technological landscape also played a role in influencing the direction of AI research. Significant limitations in computational power, storage, and available data made symbolic methods more viable and attractive compared with data- and compute-intensive connectionist methods. As symbolic systems exhausted the suite of limited problems for which they were best suited, they failed to scale precisely because of the capacities that facilitated their early success.

By contrast, early supporters of connectionism believed that research in artificial intelligence should be motivated not by problems typified of efficient symbol manipulation (at which early computers excelled) but by challenges in information processing. Here, problem domain selection was conditioned less by the extant technological landscape than by what researchers considered to be a general characteristic of intelligent behavior. It was not until that landscape evolved sufficiently to accommodate the resource demands of neural network approaches that the methods so prevalent across science today \cite{duede2024oil} could gain purchase.

The fate of symbolic systems approaches to artificial intelligence and the subsequent (re)emergence of connectionist methods illustrates a more general process typical across fields whereby early entrants bet on methods most useful for solving a set of problems that readily present themselves. In early computation, the most straightforward way to cast problems was according to strict rules and logical processes that enabled efficient symbol manipulation and so the methods best suited for solving these problems were similarly cast.\footnote{It is also worth noting that early supporters of symbolic AI were committed to the truth of the hypotheses that being a physical symbol system is required for general intelligence and that any physical symbol system can be organized in such a way as to exhibit general intelligence \cite{newell2007computer}.}

As scientists adopt the questions, methods, data, and norms of evaluation that characterize the field \cite{koch2021reduced}, research choices become increasingly conservative, focusing on well-known and highly trafficked problem areas. To a large degree, this approach is initially efficient for their careers \cite{rzhetsky2015choosing}. Over time, however, this process leads to the emergence of social contracts within fields such that signaling membership requires a focus on the field's core tenets, ultimately leading to the crystallization of well-understood field boundaries \cite{chu2021slowed}. 

Removing such boundaries often requires an antecedent change in the structure of the field itself. For instance, it has been observed that the premature death of eminent (`star') life scientists significantly impacts research behavior in their fields, leading to a decrease in contributions from long-time collaborators of the star and a notable increase in contributions from non-collaborators. This shift often results in the field evolving in new directions, as these outside contributions, which tend to draw upon different scientific approaches, represent fresh perspectives and attract disproportionate attention as captured in high citations \cite{azoulay2019does}.

These dynamics mirror processes involved in the speciation of genetic populations, leading to analogous outcomes. If populations are isolated for extended periods, cross-population genetic exchange becomes increasingly unfeasible. Conversely, if species remain in close, continuous contact, the benefits of heterosis or `hybrid vigor' are lost due to the minimal introduction of novel genetic material \cite{chen2010molecular}. Robust genetic diversity, which is crucial for evolution and resilience to environmental change, is enhanced by gene variety within single populations. This is extended with multiple, largely separate populations, each adapting to distinct environmental challenges. Such systems not only expand existing diversity but also develop new characteristics, which become increasingly significant across the population boundaries as environmental conditions change.

Returning to scientific populations, as fields homogenize and increasingly focus on circumscribed sets of methods and problems, scientists develop intuitions about the replicability and applicability of their work. For instance, a method or solution that continues to work well in the otherwise highly controlled environment of one scientific community might give the impression of robust certainty as peers with shared assumptions, instruments, and data confirm early findings. This impression might be misleading, however, if the success of these methods is context dependent, tied to the specific assumptions, circumstances and parameters of the narrow `laboratory' environment in which they were initially developed and tested. Under these circumstances, when a specific result `replicates', it is often, in fact, these laboratory environments that become inadvertently replicated through the sharing of data, methods, as well as embodied or tacit knowledge within the researchers themselves \cite{Latour1987-pr,Knorr-Cetina2013-cr,danchev2019centralized,belikov2022prediction}.

In summary, scientists may cognitively \textit{believe} they are replicating studies or confirming findings because these theories are true or best explain their focal phenomena when, in fact, replication occurs because the studies themselves overly rely on shared techniques, assumptions, data, and even people. This sharing naturally reduces the variability and distance between such investigations, leading to greater uniformity in research outcomes and the psychological impression of confidence. Wimsatt noticed an analog of this process when he noted that the procedures used for the robust detection of some invariant target phenomenon ``require at least partial independence of the various processes across which invariance is shown. And each of them is subject to a kind of systematic error leading to a kind of illusory robustness when we are led, on less than definitive evidence, to presume independence and our presumption turns out to be incorrect'' \cite[pg.46]{wimsatt2007re}.

In this way, as a field matures and becomes more homogeneous, with researchers increasingly adopting the same assumptions, methods, and data sets, the likelihood of replication and confirmation of findings within that field increases. However, this apparent success in replication may not necessarily indicate robust or generalizable scientific certainty. Instead, it is a byproduct of reduced diversity in investigative approaches and perspectives. It is the \textit{opposite} of robust detection \cite{wimsatt1981robustness}. This shared cognitive framework can create feedback loops where the success of certain methods and assumptions reinforces their continued use, further entrenching the field's focus on a narrow range of ideas and techniques. This phenomenon can lead to an epistemic echo chamber or scientific monoculture of the kind Kuhn describes, where researchers are more likely to confirm and reinforce existing beliefs and approaches within their fields, not only at the expense of innovation but also critical examination and calibrated confidence in the broader applicability of findings.\footnote{Of course, for Kuhn, this homogenization process is required for paradigm-driven science and furnishes fields with the necessary theoretical and empirical frameworks for crises and their resolution. That is, in whatever sense science can be viewed as progressive, for Kuhn, what we have described here is required for progress.} The psychological or cognitive corollary that we highlight here implies that this is not merely a structural or systemic issue \cite{zollman2010epistemic} but also involves the perceptions and mental models of the scientists themselves.

\section*{The Puzzle of Epistemic Trust}
The preceding discussion suggests that a significant challenge for collective certainty is the homogenization of fields around similar observational data, methods, and collaborators as a means for addressing shared problems. However, in tension with this challenge, scholars have argued that such homogeneity is epistemically necessary for scientific practice precisely because it provides the conditions for both social and epistemic trust among scientists. Philosophers and social scientists largely agree that without epistemic and social trust among scientists, contemporary science would not be possible \cite{fricker2006second,frost2013moral,gerken2015epistemic}. Working in recognizable or respected labs and having acquired recognizable or respected training are prerequisites for epistemic trust in scientific settings where the process of conducting research relies upon scientists with specialized expertise. Given that individual members of a research team will likely lack the expertise necessary to understand, skillfully execute, and expertly evaluate all that is required for reaching and ratifying a result, trust is necessary for progress.

This tension presents a puzzle for contemporary science. On the one hand, epistemic and social trust are required for scientific advance. But to build social and epistemic trust, homogenization is also required, which reduces the independence required for genuine replicability (e.g., conceptual replicability), as results become a reliable output of reproduced methods and data. 

\section*{Current Developments in Science Exacerbate the Challenge}

In recent years, various large-scale initiatives have been designed and implemented with the explicit aim of overcoming the replicability challenge. Nevertheless, we argue that rather than exerting a mitigating influence, some of these initiatives have the counter-intuitive effect of exacerbating the challenge. Consider, for instance, the `Open Science' movement. Open Science attempts to resolve the replicability challenge by realizing a communal scientific system in which data, methods, and results are made openly available and rapidly communicated. At its most ambitious, for instance, making the entire scientific process `FAIR' (findable, accessible, interoperable, reusable) has become a key framework for researchers and institutions aiming to make their digital research outputs more open and usable by more scientists and machines.

Initially, the effect is highly generative, surfacing a bewildering array of approaches, results, and methods analogous to a Cambrian explosion. Naturally, however, this leads to competitive dynamics in the economy of attention. While, intuitively, this might appear expedient, the result is that the best-performing early codebases and datasets can rapidly harden into the `standard' or `state-of-the-art' moving forward, crowding out alternative approaches that are equally promising but, perhaps, more challenging to realize initially. Dynamics of this kind serve to generate conditions that look \textit{less} like the free exchange of ideas and more like \textit{competitions} that ultimately drive down the diversity of code and data for reasons independent of their epistemic merit. For example, researchers might be induced to abandon promising but hitherto fruitless paths to pursue approaches that have yielded early successes. Recall the early history of artificial intelligence research which cemented symbolic approaches for decades. Similarly, frugal funding agencies and foundations requiring Open Science approaches like open data and open code may, in the medium term, be less willing to sustain datasets and codebases that appear `redundant', `underperforming', or that fail to rapidly lock in a substantial number of users. The result is a realization of the Campbell's Law \cite{campbell1979assessing}, Goodhardt's Law \cite{goodhart1984problems}, and the Lucas Critique \cite{lucas1976econometric} suggesting that when science comes to be evaluated on fixed measures of quality, then incentives to maximize the surrogate objectives or measures drive the correlation of those measures with quality to zero.

One interesting outcome is that approaches like the Open Science movement, which aims to improve replicability, are good at generating technologies. As we argued above, the generation of and innovation upon new technologies requires successful reproduction of the technology. In fact, this dynamic has been observed in the effect of benchmarking in computer science research where the establishment of state-of-the-art performance on specific large, centralized, and open datasets, explicitly curated for FAIR science, leads to an increasingly narrow character of replication \cite{koch2021reduced}. Here, success exhibits a certain fixity to established prior successes and serves to lock in early leading approaches, thereby discouraging novel entrants. Nevertheless, as we discuss below, if the generation and maintenance of seemingly redundant and underperforming approaches is treated as a fixed cost of doing science, then open and continuing competition becomes an epistemic good precisely because it broadens and sustains search \cite{rzhetsky2015choosing}.

This problem has the potential to become particularly pernicious in `Big Science' where results depend on the development of \textit{singular}, massive, and highly expensive scientific instruments. The expense and specialization required to build and maintain such instruments causes these instruments, the questions they can resolve, and the underlying theoretical commitments that motivated and justified their construction to become entrenched \cite{wimsatt1986developmental}. As a result, Big Science comes to exhibit the same pattern we have observed again and again. By requiring a level of investigative convergence that increases tacit knowledge, entire fields can become bound to historical, backward compatible designs and struggle to pivot in the direction of new questions that might require completely novel instrumentation and assumptions to address. 

\section*{Science Policy Solutions}
\label{sec:solutions}

Insofar as several recent advances associated with science: open science; big science; and collective science may exacerbate the puzzle of collective certainty, here we explore solutions. These include designed, evolved, and imagined institutions and technologies that might curb it to enable greater and more confident scientific certainty. They include (1) the corporeal university, (2) reimagined education as experimentation, (3) competition policy for science, and (4) building AI as `alien intelligence' to step beyond our epistemic bubbles. 

\subsection*{The Corporeal University}
\label{sec:corp_uni}
The word \textit{university} in the Middle Ages referenced not only a body of schools engaged in higher instruction across the branches of ancient knowledge (e.g., the trivium and quadrivium), but the entirety of all created things. In the modern world, universities distinguish themselves from colleges and institutes by their commitment to the full spectrum of modern subjects, ranging across the humanities, social and natural sciences. Our own prior work has demonstrated that the diversity underlying universities, and the surprising conversations it routinely generates, form the most sustained pattern of innovation we find in our own study of intellectual influence across science and scholarship. In our study, we developed customized author surveys regarding the intellectual influence of referenced work on scientists' own published papers, combined with precise measures of institutional and semantic distance between focal and referenced works \cite{teplitskiy2022status}. Statistical models revealed that being at the same institution \textit{but not} the same department was strongly associated with the attribution of intellectual influence on scientists' and scholars' published work. This influence increased with intellectual distance between authors---the more different the referenced work done by colleagues at one's institution, the more influence it was to have on one's own. How do professors and graduate students discover the diverse work of others within the university? Through living together in the same physical location, which catalyzes serendipitous meetings that occur for reasons primarily other than science and scholarship--co-attendance at childrens' sporting games, service on university committees, shopping in the same supermarkets and dining at the same restaurants \cite{duede2024being}. 

This effect is further reinforced by recent work highlighting how scholars are much more likely to design and publish breakthrough research together in-place \cite{lin2023remote}. This likely stems from frictions on deep, interactive thinking online \cite{brucks2022virtual} and barriers to the discovery, connection, and recruitment of diverse researchers who have the potential to surprisingly influence one another. Universities worldwide constitute places where people doing very different work engage in sustained interactions through departments, committees, seminars, and communities \cite{duede2024being}. These interactions come to uniquely influence scientists' published research, suggesting their continuing importance for sustainable advance. This ``coordination by accident'' has primarily been an accidental byproduct of the centralization of education. Our theory suggests the potential to broaden and optimize university diversity to create more consistently fruitful connections across scientists and disciplines.

Rather than lean in this direction, the corporeal university with physically situated scholars is currently under threat. In the post-COVID world of more distanced work and with the rise of mega-online universities, like the UK's Open University and Arizona State University with more than 100,000 students each engaged in distanced classes with distanced faculties, many universities are increasingly out-of-place. Moreover, with European efforts to create ``centers of excellence'' that bring together scholars of one or a few connected disciplines, the most routine connections are maximized and the most surprising ones minimized. We call for a rethinking of universities, sustained conferences and workshops that explicitly design diversity for novel discovery and deep verification of existing knowledge. 

\subsection*{Education as Experiment}
University education is largely controlled by disciplinary departments. As we argued above, disciplines represent commitments to a problem and the methods expected to best solve it. In this way, disciplines encode a \textit{bet}, often made long ago, on the efficacy of an approach to a valued challenge. These become enshrined as tradition in disciplinary education (e.g., the required ``methods sequence''), and later as the epistemic standard for knowledge on that topic. Mathematics is advanced by proof; cultural theory by participant observation; social psychology by experimental mechanism discrimination; and paleontology by fieldwork and specimen comparison. The problem follows our discussion earlier. New cases are likely to yield diminishing marginal returns from the sanctioned approach. When scientists and scholars forge a new method, they strategically select the best initial case on which to demonstrate it. Insofar as they are prescient, the method will wane rapidly in utility for new cases and new problems. 

We propose re-imaginging education as an exploration, where every student (or class) is an experiment to explore the value of a new synthesis of methods and problems \cite{shi2023surprising}, mirroring the thinking and learning strategies of very young children \cite{gopnik2012scientific}. This would require both college and graduate education to be owned and administered by the university, or shifting collections of departments, and not specific disciplines. This reflects efforts by national funding bodies like the U.S. National Science Foundation and National Institutes of Health to sponsor multi-disciplinary educational programs, often linking emerging methods like artificial intelligence with an established problem areas like structural biology on which traditional methods had stalled \cite{bryant2022improved}. Such an approach raises ethical questions. By violating the staffing model of education that seeks to endow students with labor market signals, ``experimental students'' may fail to achieve synergy or yield an interpretable brand. Nevertheless, students and not faculty are those most motivated to invest in a novel synthesis of problem and solution, and if scientific education is to realize its stated goal of enabling outsized discovery, it must also allow for failure, and potentially compensate for it with an expanded safety net. In short, we propose that, rather than pursue an exhausted bet of methodological utility, instead we build a portfolio of new possibilities, supported by the cultivation of new fields for disciplinary cross-breeding and interaction. Such an educational approach will not only deepen the search for useful knowledge, but it will also broaden the approaches through which established knowledge is assayed for continuing relevance.   

\subsection*{Competition Policy for Science}
In the United States and other market regimes, competition is preserved within markets by legislative, executive, and judicial policies to maintain the diversity of ownership and interests in the service of fostering continued innovation. Monopolies occur when only a singular producer of a good or service exists within a market. The singularity of motivation and interest that results allows the monopolist to restrict production, raise price, and suppress innovation \cite{akcigit2023connecting, akcigit2023happened}. As demonstrated above, when singular institutions or networks take control of a particular scientific theory or problem, similar dynamics occur. They may raise epistemic barriers and decrease competition by limiting the diversity of approaches that can be published in top journals or receive funding from centralized funding bodies \cite{azoulay2019does}. When star scientists die, innovation only emerges when their former students do not control top journals and grant review panels. 

Like the U.S. Securities and Exchange Commission (SEC), which prohibits company mergers that risk decreasing competition and innovation within the economy, we propose the development of a capacity to evaluate and advise on issues of competition policy in research. This could be a new, centralized, research agency, the distributed capacity to evaluate funding within science-funding agencies, or a research network that provides \textit{ad hoc} guidance. Prominent funders like the U.S. National Institutes of Health and the National Science Foundation often centralize funding around large-scale centers and initiatives. They also have mechanisms to provide renewals for funded projects that validate and extend prior work. Competition policy would provide guidelines and strategies that enable scientific funders to build epistemic, cognitive, and empirical (e.g., data, methods) diversity into and across funding programs to ensure that: (1) important facts are subjected to independent assessment; and (2) critical research initiatives are approached from multiple directions and perspectives. 

The existence of this capacity reflects some aspects of the speculative funding model present in the U.S. Defense Advanced Research Projects Agency (DARPA) and its growing number of copcats in other areas (e.g., Intelligence-ARPA; ARPA-Energy; ARPA-Health; etc.) DARPA seeks to pull together diverse approaches to solve problems or validate critical ideas, then ``fail fast'' by ruthlessly pruning all but the winning directions. 

The motivation for competition policy in research is very similar to that in business. We seek to maximize sustained competition between approaches in science to validate a robust and dynamically updated fact-base on which future work can reliably build. Such an approach would also catalyze novel discovery by keeping alternative approaches alive and in contact in the service of sustained advance. Such an approach may be considered an unwelcome contraint by some who seek to sponsor their preferred scientific tribe at the cost of others. Nevertheless, research funding represents an increasing percentage of federal government budgets, and as we have sought to demonstrate here, competition matters critically for both scientific discovery and validation. 

\subsection*{AI as Alien Intelligence}
Artificial intelligence (AI) is emerging as a powerful general-purpose technology in society, especially in the era of Large Language Models (LLMs), which are natively programmed with human speech and text. In research areas ranging from materials discovery and drug development to high-energy physics and molecular biology, AI has intervened in processes of scientific search, validation, hypothesis generation, and discovery \cite{duede2023deep}. AI for science often relies on published findings and observed or experimental data but typically ignores the distribution of scientists and inventors \cite{teruya2022arts}, the human prediction engines who continuously alter the landscape of discovery and invention. We argue that reconceptualizing AI as alien intelligence could allow us to expand and accelerate the diversity of scientific approaches for both scientific validation and discovery.

Consider the potential to create both cultural and cognitive alien intelligence. 
In prior work, we incorporated the knowledge of human researchers to improve predictions of future discoveries compared with AI methods that ignore them. Then we used these simulated human scientists to generate AI methods that avoided the human crowd and identified scientific possibilities that, without AI, would have remained unimagined and unexplored until the more distant future \cite{rzhetsky2015choosing}. By identifying and correcting for collective patterns of human attention, formed by field boundaries and institutionalized education, these `alien' intelligence models complemented the contemporary scientific community \cite{sourati2023accelerating}. The hypotheses our AI advanced made science more cognitively diverse by leveraging combinations of experience and literature inaccessible to any current human scientist, but they were not incomprehensible to them. In this way, such AI represents a `cultural alien', unfamiliar but understandable to human scientists. Alternatively, one can imagine constructing `cognitive' alien intelligence predictions that cannot be understood or represented by a human scientist. Such a prediction might have greater than human cognitive complexity, like a 1000-dimensional manifold of scientific associations, which limits its ability for flexible communication and generalization. 

Despite these possibilities, current efforts to incorporate AI into science often rely on an increasingly mono-cultured field of available AI architectures, typically built around deep neural networks and, more specifically, the transformer architecture underlying LLMs \cite{kleinberg2021algorithmic}. If students from a high school English class all use ChatGPT, currently the most popular LLM, to respond to the same prompt, their essays will become more similar. One possible future of AI-enhanced science involves a similar narrowing of AI approach, applied to many scientific tasks. Armed with awareness of the importance of scientific diversity in processes of validation, abduction, and discovery, scientists and engineers can work to diversify the AI architectures used to validate and extend scientific discovery.  

\section*{Conclusion}

In this chapter, we developed and explored an epistemological paradox in science, the puzzle of collective certainty. This puzzle involves how the more scientists work together in research, the less their collaborative work contributes to the value of collective certainty \cite{belikov2022prediction, danchev2019centralized}. As scientists share problems, data, methods, and collaborators, it increases the positive reinforcement they receive for shared beliefs as they become less independent of those around them \cite{kang2023limited,sourati2022data}. This raises concerns that many scientists reside in epistemic ``bubbles'' of their own making, which appear robust inside these insular communities, but do not extend beyond. This is paradoxical because being aware of another's expertise is a prerequisite for epistemic trust, but this also increases the likelihood that those evaluating and evaluated will come to share the same methods and assumptions \cite{shapin1995social,fricker2006second,frost2013moral,gerken2015epistemic} regardless of validity. 

The validity of scientific results implies conceptual replication. This enables the identification of phenomena robustness---its ability to be multiply detected and realized across a range of procedures. On this view, scientific results become valid as they obtain in all nearby ``worlds'', where distinct methods deployed by distinct scientists on distinct data reveal the same phenomena. While all valid results are, in principle, reproducible and replicable, results that are \textit{merely} replicable cannot, in general, constitute useful science if scope conditions are narrow or under-specified. In this chapter, we first reviewed the widely popularized ``replication crisis'' in science, then explored a puzzle in the dynamics of collective certainty. Next, we examined developments in science that, while seeking to improve conceptual replicability, in fact exercerbate the challenge. Finally, we proposed four possible science policy solutions that might counter and correct the challenge.

We call upon both natural and social science to consider measures that allow them to monitor, increase, and sustain epistemic diversity across scientific applications. Moreover, we enjoin science policy to design and manage diverse ecologies of approach that allow epistemic evolution and diversification. In Adam Smith's \textit{Wealth of Nations}, he proposed a weak competition policy that prohibited manufacturers from meeting or formulating a public registry to avoid collusion. We argue for the opposite, that measuring dependencies in science is the first step to regulating and reducing them on the path to firmer facts and more surprising discoveries. 

\bibliographystyle{alpha}
\bibliography{main}
 
\end{document}